  \documentclass[12pt]{article}
  \usepackage{amssymb}
  \usepackage{latexsym}
  \usepackage{graphicx}

 \catcode`\@=11 \@addtoreset{equation}{section}\catcode`\@=12

\newcommand{\R}{ {\mathbb R} }
\newcommand{\fnm}{\footnotemark}
\newcommand{\fnt}{\footnotetext}

\begin{document}

 \begin{center}

 \large \bf On stability of exponential cosmological   solutions
with non-static volume factor in  the Einstein-Gauss-Bonnet  model
  \end{center}

 \vspace{0.3truecm}

 \begin{center}

  V. D. Ivashchuk  \fnm[1]\fnt[1]{e-mail:  ivashchuk@mail.ru}

\vspace{0.3truecm}

 \it Center for Gravitation and Fundamental Metrology,
 VNIIMS, 46 Ozyornaya ul., Moscow 119361, Russia

 \it Institute of Gravitation and Cosmology,
 Peoples' Friendship University of Russia
 (RUDN University), 6 Miklukho-Maklaya ul.,
 Moscow 117198, Russia

\end{center}

 \begin{abstract}

 A $(n+1)$-dimensional  gravitational model with Gauss-Bonnet term and cosmological constant term  is  considered. When  ansatz with diagonal  cosmological  metrics is adopted,  the solutions  with  exponential dependence of scale factors:  $a_i \sim \exp{ ( v^i t) }$, $i =1, \dots, n $, are analysed for $n > 3$.
 We study the stability of the solutions  with non-static volume factor, i.e. if
 $K(v) =  \sum_{k = 1}^{n} v^k \neq 0$.  We prove that under certain restriction $R$ imposed  solutions with
  $K(v) > 0$ are stable  while  solutions with $K(v) < 0$ are unstable. Certain examples of  stable solutions are presented. We show that the  solutions with   $v^1 = v^2 =v^3 = H > 0$ and  zero variation of the effective gravitational constant  are stable if the restriction $R$ is obeyed.

  \end{abstract}

\clearpage

\section{Introduction}

This paper is devoted to $D$-dimensional gravitational model with the
so-called Gauss-Bonnet term. It is governed by the action
\begin{equation}
 S =  \int_{M} d^{D}z \sqrt{|g|} \{ \alpha_1 (R[g] - 2 \Lambda ) +
             \alpha_2  {\cal L}_2[g] \},
   \label{1.1}
 \end{equation}
where $g = g_{MN} dz^{M} \otimes dz^{N}$ is the metric defined on
the manifold $M$, ${\dim M} = D$, $|g| = |\det (g_{MN})|$
and
\begin{equation}
  {\cal L}_2 = R_{MNPQ} R^{MNPQ} - 4 R_{MN} R^{MN} +R^2
   \label{1.2}
 \end{equation}
is the quadratic  ``Gauss-Bonnet term'' and $\Lambda$ is cosmological term.
Here $\alpha_1$ and $\alpha_2$ are non-zero constants.
The appearance of the  Gauss-Bonnet term  was motivated by string theory
\cite{Zwiebach,GW,MTs}.

 At present, the so-called Einstein-Gauss-Bonnet (EGB) gravitational model which is governed
 by the action (\ref{1.1}) and  its modifications are intensively used in  cosmology, see
 \cite{Ishihara} - \cite{EIK}  and references therein,  e.g. for explanation  of  accelerating
 expansion of the Universe following from  supernovae (type Ia) observational data \cite{Riess,Perl,Kowalski}.
 
 Here we consider  the cosmological  solutions with diagonal
 metrics governed by $n$ scale factors  depending upon one variable, where $n > 3$; $D = n + 1$.
 We study the stability of  solutions with exponential dependence of scale factors
 with respect to the synchronous time variable $t$
  \begin{equation}
  \label{1.3}
  a_i(t) \sim \exp{(v^i t )},
  \end{equation}
   $i = 1, \dots, n$.
  In our analysis  we restrict ourselves by  a  class of
  perturbations which depend on $t$ and do not disturb the  diagonal form of the metric.

  For possible physical applications
  solutions  describing   an exponential isotropic expansion of  3-dimensional flat factor-space, i.e.   with
  \begin{equation}
  \label{1.4}
  v^1 = v^2 =v^3 = H > 0,
  \end{equation}
   and small enough variation of the effective gravitational constant $G$  are of interest.
 We remind that $G$ (for $4d$ metric in Jordan frame, see Remark 4 in Section 4) 
 is proportional to the inverse volume scale factor
 of the internal space, see  \cite{BIM,Mel,IvMel-14} and refs. therein.
 Due to experimental data, the variation of $G$ is allowed
 at the level of $10^{-13}$ per year and less.   The  most stringent limitation  on $G$-dot
 (coming from the set of ephemerides)    was obtained in ref.  \cite{Pitjeva}
   \begin{equation}
     \label{1.G2}
    \dot{G}/G = (0.16 \pm 0.6) \cdot 10^{-13} \ {\rm year}^{-1}
    \end{equation}
    allowed at 95\% confidence (2-$\sigma$).

 Here we reduce the set of cosmological equations
 to the (mixed) set of algebraic and differential equations
  \begin{eqnarray}
      f_0(h) = 0,
     \label{1.5}   \\
      f_i(\dot{h},h) =0.
      \label{1.6}
     \end{eqnarray}
 where $h = h(t) = (h^i(t) = \dot{a}_i(t)/ a_i(t))$ is the set of so-called  ``Hubble-like''   parameters corresponding to scale factors $a_i(t)$;  $f_0(h)$ and $f_i(\dot{h},h)$ are polynomials of the fourth order in $h^i$;  $f_i(\dot{h},h)$ are polynomials of the first order in $\dot{h}^i$.
  The fixed point solutions  $h^i(t) = v^i$ ($i = 1, \dots, n$)
  correspond to exponential solutions of   (\ref{1.3}). They  obey a set of
  $n+1$ polynomial equations of the fourth order.
  We analyze the stability of the  fixed point solutions by imposing the following restriction
  \begin{equation}
    \label{1.7}
    ({\rm R }) \quad  \det \left( \frac{\partial f_i}{\partial \dot{h}^j}({\bf 0},v) \right) \neq 0,
    \end{equation}
 which guarantees the local resolution of eqs.  (\ref{1.6}) in the vicinity of the point
 $({\bf 0}, v) \in \R^{2n}$: $ \dot{h}^i = \varphi^i(h)$ with $\varphi^i(v) = 0$,
 $i = 1, \dots, n$. Here ${\bf 0} = (0, \dots, 0) \in \R^{n}$.

  We also impose another restriction on $v$:
 \begin{equation}
   \label{1.8}
   \sum_{k = 1}^{n} v^k \neq 0,
   \end{equation}
  which means that the solutions with constant volume scale factor are not considered here.
  We note that a solution with $\sum_{k = 1}^{n} v^k = 0$ obeying
  (\ref{1.4})  gives an enormously big value of the variation of $G$:  $\dot{G}/G =  3 H$,
    where $H$ is  the Hubble parameter, see Remark 5 in Section 4 below.
    This value of $G$-dot contradicts to  the observational restrictions, e.g. (\ref{1.G2}).
     We remind that the present value of $H$ is
    $(6.929  \pm 0.157) \cdot 10^{-11} \ {\rm year}^{-1}$  \cite{Ade} (with 95\% confidence level).

  The main result of the paper is the following one:
 fixed point solutions $h(t) = v$ to eqs. (\ref{1.5}) and (\ref{1.6}),
 which obey restrictions (\ref{1.7}) and (\ref{1.8}), are stable if
 and only if  $\sum_{k = 1}^{n} v^k  > 0$.  This result is in agreement with
  the approach of S. Pavluchenko from ref. \cite{Pavl-15},
  see Remark 2 in Section 3 below.

The paper is organized as follows. In Section 2
the equations of motion for  $D$-dimensional EGB model  are considered.
For  diagonal  cosmological  metrics the equations of motion are equivalent
to a set of Lagrange equations corresponding to a certain ``effective'' Lagrangian.
The Lagrange equtions for a certain choice of the lapse function (corresponding
to the synchronous time variable) are reduced to the set of eqs. (\ref{1.5}), (\ref{1.6}).
Section 3 is devoted to analysis of stability of the exponential solutions with constant Hubble-like
parameters: here a set of equations for
perturbations $\delta h^i(t)$ (obtained in linear approximation) is studied and  general solution
to these equations is found. The main proposition on stability of exponential solutions (Proposition 2)
is proved. In Section  4  some examples of  stable cosmological solutions with exponential behavior of scale factors  are presented.

\section{The   model}

 \subsection{The set-up}

 Here we consider the manifold
 \begin{equation}
   M = (t_{-},t_{+})  \times  M_1 \times \ldots \times M_n , \label{2.1}
 \end{equation}
  with the metric
 \begin{equation}
  g= - e^{2{\gamma}(t)} dt \otimes dt  +
 \sum_{i=1}^{n} e^{2\beta^i(t)}  dy^i \otimes dy^i,
 \label{2.2}
 \end{equation}
 where   $i = 1, \dots, n$;
  $M_1, ...,  M_n$  are one-dimensional manifolds (either $\R$ or $S^1$)
  and $n > 3$.
  The functions ${\gamma}(t)$ and
 $\beta^i(t)$,  $i = 1,\ldots, n$, are smooth on
 $(t_{-},t_{+})$.

 For physical applications  we put $M_1 =M_2 = M_3 = \R$, while
 $M_4, ..., M_n$ may be considered to be compact ones (i.e. coinciding with
 $S^1$).

 The integrand in  (\ref{1.1}), when the
 metric (\ref{2.2}) is substituted, reads as follows
   \begin{equation}
    \sqrt{|g|} \{ \alpha_1 R[g] +
               \alpha_2  {\cal L}_2[g] \} = L + \frac{df}{dt},
    \label{2.3}
   \end{equation}
 where
    \begin{equation}
        \label{2.4}
    L = \alpha_1 (e^{-\gamma + \gamma_0} G_{ij}
        \dot{\beta}^i \dot{\beta}^j - 2 \Lambda e^{\gamma + \gamma_0 })
      - \frac{1}{3} \alpha_2  e^{- 3 \gamma + \gamma_0}
        G_{ijkl} \dot{\beta}^i \dot{\beta}^j \dot{\beta}^k  \dot{\beta}^l,
   \end{equation}
   $\gamma_0 = \sum_{i =1}^{n} \beta^i$ and
  \begin{eqnarray}
       G_{ij} = \delta_{ij} -1,
         \label{2.10}   \\
       G_{ijkl}  = G_{ij} G_{ik} G_{il} G_{jk} G_{jl} G_{kl}
       \label{2.11}
      \end{eqnarray}
      are respectively the components of two  metrics on
      $\R^{n}$ \cite{Iv-09,Iv-10}. The first one is ``minisupermetric'' - 2-metric
      of  pseudo-Euclidean signature  and the second one is the Finslerian 4-metric
      \cite{Iv-09,Iv-10}.
     Here we denote $\dot{A} = dA/dt$ etc. The function $f(t)$ in (\ref{2.3}) is irrelevant
     for our consideration (see \cite{Iv-09,Iv-10}).

      In derivation of (\ref{2.4})
        the following identities \cite{Iv-09,Iv-10} were used:
      \begin{eqnarray}
       G_{ij}v^i v^j = \sum_{i =1}^{n} (v^i)^2 -
        (\sum_{i =1}^{n} v^i )^2 = S_2 - S_1^2,
         \label{2.12}   \\
       G_{ijkl}v^i v^j v^k v^l  = S_1^4- 6 S_1^2 S_2
         +   3 S_2^2 + 8  S_1 S_3  - 6 S_4.
       \label{2.13}
      \end{eqnarray}
 Here and in what follows $S_k = S_k(v) = \sum_{i =1}^n (v^i)^k$.

      The definitions (\ref{2.10}) and (\ref{2.11}) imply
      \begin{eqnarray}
       G_{ij}v^i v^j = -2 \sum_{i < j} v^i v^j,
         \label{2.14}   \\
       G_{ijkl}v^i v^j v^k v^l  = 24 \sum_{i < j < k < l} v^i v^j v^k v^l.
       \label{2.15}
      \end{eqnarray}

  The equations of motion corresponding to the action (\ref{1.1})
  have the following form
  \begin{equation}
   {\cal E}_{MN} = \alpha_1 {\cal E}^{(1)}_{MN} + \alpha_2 {\cal E}^{(2)}_{MN} = 0,
   \label{2.3e}
 \end{equation}
  where
  \begin{eqnarray}
   {\cal E}^{(1)}_{MN} = R_{MN} - \frac{1}{2} R g_{MN} + \Lambda g_{MN},
   \label{2.3a} \\
   {\cal E}^{(2)}_{MN} = 2(R_{MPQS}R_N^{\ \ PQS} - 2 R_{MP} R_N^{\ \ P}
   \nonumber \\
   -2 R_{MPNQ} R^{PQ} + R R_{MN}) -  \frac{1}{2} {\cal L}_2  g_{MN}.
   \label{2.3b}
   \end{eqnarray}

    It may be  shown (along a line as it was done in
    \cite{Iv-10} for the case $\Lambda = 0$) that the field eqs. (\ref{2.3e}) for the metric
    (\ref{2.2}) are equivalent to the Lagrange equations
    corresponding to the Lagrangian $L$ from (\ref{2.4}).

    Thus, eqs. (\ref{2.3e}) read as follows
    \begin{eqnarray}
      \alpha_1   (G_{ij} \dot{\beta}^i \dot{\beta}^j + 2 \Lambda e^{2 \gamma })
        -  \alpha_2  e^{- 2 \gamma}
         G_{ijkl} \dot{\beta}^i \dot{\beta}^j \dot{\beta}^k
         \dot{\beta}^l = 0,  \label{2.17} \\
         \frac{d}{dt}[ 2 \alpha_1  G_{ij} e^{-\gamma +
         \gamma_0}
          \dot{\beta}^j
         - \frac{4}{3} \alpha_2 e^{- 3 \gamma + \gamma_0}
         G_{ijkl}  \dot{\beta}^j \dot{\beta}^k
         \dot{\beta}^l] - L = 0,   \label{2.18}
      \end{eqnarray}
     $i = 1,\ldots, n$; and $L$ is defined in  (\ref{2.4}).

  Now we put $\gamma = 0$. By introducing ``Hubble-like'' variables $h^i = \dot{\beta}^i$,
   eqs. (\ref{2.17}) and  (\ref{2.18}) may be rewritten as  follows

  \begin{eqnarray}
     E = E(h) \equiv G_{ij} h^i h^j + 2 \Lambda  - \alpha G_{ijkl} h^i h^j h^k h^l = 0,
         \label{3.1} \\
     U_i = U_i(\dot{h}, h) \equiv \frac{d L_i}{dt}  +  (\sum_{j=1}^n h^j) L_i -  L_0 = 0,
                \label{3.2}
      \end{eqnarray}
      where $\alpha = \alpha_1/\alpha_2$,
      \begin{equation}
       L_0 =  G_{ij} h^i h^j - 2 \Lambda
              - \frac{1}{3} \alpha G_{ijkl} h^i h^j h^k h^l,
      \label{3.2L}
      \end{equation}
      and
      \begin{equation}
       L_i = L_i(h) = 2  G_{ij} h^j
       - \frac{4}{3} \alpha  G_{ijkl}  h^j h^k h^l  \label{3.3},
      \end{equation}
     $i = 1,\ldots, n$.
      Thus,  we are led to the   autonomous system of the first-order differential equations on
      $h^1(t), \ldots, h^n(t)$  (see \cite{Iv-09,Iv-10} for $\Lambda = 0$).

       Due to (\ref{3.1}) we have
           \begin{equation}
            L_0 =   \frac{2}{3} (G_{ij} h^i h^j -  4 \Lambda).
             \label{3.4}
           \end{equation}

  In what follows we will use instead of (\ref{3.1}), (\ref{3.2})
  an equivalent set of equations:  (\ref{3.1}) and
      \begin{equation}
        Y_i = Y_i(\dot{h}, h) \equiv \frac{d L_i}{dt}  +  (\sum_{j=1}^n h^j) L_i -
        \frac{2}{3} (G_{ij} h^i h^j -  4 \Lambda) = 0.
            \label{3.2a}
        \end{equation}
  We note that the following identity is valid
  \begin{equation}
      U_i(\dot{h}, h) =  Y_i(\dot{h}, h) - \frac{1}{3} E(h),
       \label{3.2b}
  \end{equation}
   $i = 1,\ldots, n$.

   Eqs. (\ref{3.1}) and  (\ref{3.2a}) are dependent, since
    \begin{equation}
    h^i Y_i = \frac{dE}{dt} + \frac{4}{3} (\sum_{j=1}^n h^j) E.
                \label{3.5a}
    \end{equation}
    This identity may be proved by using two relations:
     \begin{eqnarray}
     h^i \frac{dL_i}{dt} = \frac{dE}{dt},
          \label{3.6a} \\
      h^i L_i  =  L_0  +  \frac{4}{3} E,
          \label{3.7a}
     \end{eqnarray}
  following  from (\ref{3.1}) and (\ref{3.3}).

   \subsection{Useful relations}

         In what follows we use the  definitions
           \begin{equation}
            B = B(v) = G_{ijks} v^i v^j v^k v^s,
            \qquad A_i = A_i(v) = G_{ijkl} v^j v^k v^l.
              \label{3.9}
           \end{equation}

         For isotropic case
            \begin{equation}
             v = (v^i) = (H,\ldots,H)
             \label{3.11}
              \end{equation}
         we get
           \begin{equation}
           B = n(n-1)(n-2)(n-3) H^4, \qquad  A_i = (n-1)(n-2)(n-3)H^3,
                  \label{3.12}
          \end{equation}
         $i = 1,\ldots, n$.

         Here we deal with the  ansatz which contain  two Hubble parameters
         \begin{equation}
                   v = (v^i) = (H,\ldots,H,h,\ldots,h)
                   \label{3.13}
          \end{equation}
         where $H$ appears $m$-times and $h$ appears $l$-times, $n = m + l$.
         In what follows we adopt the following agreement for indices:
         $\mu, \nu, \dots = 1, \dots, m$; $\alpha, \beta, \dots = m + 1, \dots, n$.
         Thus, $v^{\mu} = H$ and $v^{\alpha} = h$.

         We obtain
          \begin{eqnarray}
           B = m (m-1)(m-2)(m-3) H^4 + 4 m (m-1) (m-2) l H^3 h +
           \nonumber \\
            6 m (m-1)l(l- 1)H^2 h^2 + 4 m l (l - 1)(l - 2) H h^3
                                    + l (l-1)(l-2)(l-3) h^4
                \label{3.13a}
           \end{eqnarray}
           and
             \begin{eqnarray}
              A_{H} \equiv A_{\mu} = (m-1) (m-2)(m-3)H^3 + 3(m-1)(m-2)lH^2 h
                    \nonumber \\
                  + 3 (m-1)l(l- 1)H h^2 +l(l-1)(l-2)h^3,
                                 \label{3.14H} \\
              A_{h} \equiv A_{\alpha} = m(m-1)(m-2)H^3 + 3 m (m-1)(l-1)H^2 h
                    \nonumber \\
                  +3 m (l-1)(l-2)H h^2 +(l-1)(l-2)(l-3)h^3 .
                   \label{3.14h}
                 \end{eqnarray}

              We also denote
                  \begin{equation}
                   S_{ij}= G_{ijks} v^k v^s,
                   \label{3.15}
                  \end{equation}
              We note that $S_{ij} = S_{ji}$ and $S_{ii} = 0$.
                 For isotropic case (\ref{3.11}) we obtain
                 \begin{equation}
                   S_{ij}= (n -2)(n-3)H^2, \qquad i \neq j.
                       \label{3.16}
                \end{equation}

                For the  the  ansatz   (\ref{3.13})  we obtain
                 \begin{eqnarray}
                 S_{HH} =  (m-2)(m -3) H^2+ 2(m-2)lHh + l(l - 1)h^2 ,
                                \label{3.17HH}   \\
                 S_{Hh} = (m-1)(m -2)H^2 + 2(m-1)(l - 1)H h + (l- 1)(l- 2)h^2,
                                                  \label{3.17Hh}   \\
                 S_{hh} = m(m-1)H^2 + 2m(l - 2)Hh+ (l- 2)(l- 3)h^2.
                                  \label{3.17hh}
                 \end{eqnarray}
                 Here we denote:  $S_{\mu \nu}  = S_{HH}$ for  $\mu \neq \nu$;
                 $S_{\mu \alpha} = S_{\alpha \mu } = S_{Hh}$;   $S_{\alpha \beta}  = S_{hh}$
                 for $\alpha \neq \beta$.

\subsection{Polynomial equations for solutions with constant  $h^i$  }

    Let us consider the following solutions to
     eqs.  (\ref{3.1}) and (\ref{3.2a})
       \begin{equation}
             h^i (t) = v^i,  \label{3.4v}
       \end{equation}
       with constant $v^i$,   which correspond to the solutions
      \begin{equation}
      \beta^i = v^i t + \beta^i_0,  \label{3.4a}
      \end{equation}
       where $\beta^i_0$ are constants, $i = 1,\ldots, n$.

      In this case we obtain the metric (\ref{2.2})
      with the exponential dependence of scale  factors
      \begin{equation}
        g= - d t \otimes d t  +
        \sum_{i=1}^{n}  B_i^2 e^{2v^i t} dy^i \otimes dy^i,
        \label{3.4m}
        \end{equation}
       where  $B_i > 0$ are arbitrary constants.

      For the fixed point  $v = (v^i)$ we have the set of  polynomial equations
          \begin{eqnarray}
        E = E(v) =  G_{ij} v^i v^j + 2 \Lambda
          - \alpha   G_{ijkl} v^i v^j v^k v^l = 0,  \label{3.5} \\
        Y_i = Y_i({\bf 0}, v) = (\sum_{j=1}^n v^j) L_i(v)
              - \frac{2}{3} G_{kj} v^k v^j + \frac{8}{3} \Lambda  = 0,
             \label{3.6}
         \end{eqnarray}
  where $L_i$ is defined in (\ref{3.3}), $i = 1,\ldots, n$.
  For $n > 3$ this is the set of forth-order polynomial equations.

  Here and in what follows  we  use relations (\ref{2.12}), (\ref{2.13})
  and the following formulas
           \begin{eqnarray}
            v_i = G_{ij}v^j = v^i - S_1,
               \label{3.7}   \\
           A_i =  G_{ijkl} v^j v^k v^l
             = S_1^3  + 2 S_3 -3 S_1 S_2
             \nonumber \\
              +  3 (S_2  - S_1^2)  v^i
               +  6 S_1 (v^i)^2 - 6(v^i)^3,
             \label{3.8}
            \end{eqnarray}
           $i = 1,\ldots, n$  ($S_k = \sum_{i =1}^n (v^i)^k$).

      {\bf Proposition 1.}
    {\em For any solution $v = (v^1,\dots,v^n)$  to  polynomial eqs. (\ref{3.5}),
   (\ref{3.6}) with $n > 3$ there are no more than  three different  numbers among
   $v^1,...,v^n$, if  $\sum_{i=1}^n v^i  \neq 0$.}

    {\bf Proof.}
    Let us suppose that there exists a non-trivial solution
    $v = (v^1,\dots,v^n)$ with more than
    three different  numbers among  $v^1,\dots,v^n$.
    Due to  (\ref{3.8}), (\ref{3.6}) and  $\sum_{i=1}^n v^i  \neq 0$
    we get
    $C_0 + C_1 v^i + C_2 (v^i)^2  + C_3(v^i)^3 = 0$, $i = 1,\ldots, n$,
    with some real numbers $C_0$, $C_1$, $C_2$ and $C_3 \neq 0$.
    Let us consider the cubic equation
     $C_0 + C_1 x + C_2 x^2  + C_3 x^3 = 0$.
     Any number   $v^i$ obeys this  equation
    and hence at most three numbers among $v^i$  may be different.
    Thus, we are led to a contradiction. The proposition  is proved.
    The case $\Lambda = 0$ was considered earlier in \cite{Iv-09,Iv-10}.

{\bf Remark 1}. {\em  In pure Einsten case  ($\alpha = 0$) with $\Lambda > 0$
 we get  two exponential solutions with $v^1 = \dots = v^n = H$
and  $n(n- 1) H^2 = 2  \Lambda > 0$;  solution with $H > 0$ is an attractor for cosmological
 solutions with diagonal metrics,  as $t \to + \infty$, see \cite{IM-94} and  \cite{BIMZ} (for $\varphi = 0$).
Thus in this case   ($\alpha = 0$) we have an isotropisation
for $t \to + \infty$, while for  $t \to + 0$ we have Kasner-like behaviour of scale factors
near the singularity: $a_i(t) \sim  t^{p_i}$ with  Kasner parameters $p_1, \dots, p_n$
obeying: $\sum_{i =1}^{n} p_i  =   \sum_{i =1}^{n} p_i^2 = 1$.
In the case of EGB model with  $\Lambda$-term   we have for certain   
$\Lambda$ and $\alpha$  isotropic exponential solutions with $v^1 = \dots = v^n = H$ (see Section 4 below), 
but we also may have  partially anisotropic (PA) solutions, which obey $\sum_{i =1}^{n} v^i  \neq 0$, with:  
$v = (H,\dots, H, h,\dots,h)$ or $v = (H,\dots, H, h,\dots,h, z,\dots,z)$,
 and also solutions with $\sum_{i =1}^{n} v^i  = 0$ may take place.
 For  $\sum_{i =1}^{n} v^i  = 0$ (and certain   $\Lambda$ and $\alpha$) one may obtain  examples of totally anisotropic exponential solutions with  non-coinciding parameters among  $v^1, \dots , v^n$. 
Some of the exponential PA solutions are stable (see below) and they are attractors
 of certain subclasses of general solutions. The appearance  of   three (or less) independent scale factors in the model under consideration is  a feature of   exponential  (e.g. attractor) solutions, when restriction $\sum_{i =1}^{n} v^i  \neq 0$  is imposed.  We also note that the  metric (\ref{3.4m}) may be analyzed on symmetries (apparent or hidden)  by using the results of  ref. \cite{I-98}, i.e. Killing vectors,  isometry group, coset structure  $G/H$ etc,  may be presented.  The Proposition 2 may be also generalized to the Lovelock case \cite{Lov} --
  that  may be  a subject of  a separate publication.}

Now let us consider the ansatz (\ref{3.13}) with two Hubble parameters $H$ and $h$  with two restrictions imposed
   \begin{equation}
   m H + lh \neq 0, \qquad  H \neq h.
   \label{3.18}
   \end{equation}

   In this  case the set of $n+1$ eqs. (\ref{3.1}), (\ref{3.2})
   is equivalent to the set of three equations
    \begin{equation}
     E =0, \qquad   Y_H = 0, \qquad  Y_h = 0,
       \label{3.19}
    \end{equation}
   where  $Y_{H} = Y_{\mu}$, $Y_{h} = Y_{\alpha}$
  ($\mu = 1, \dots, m$, $\alpha = m + 1, \dots, n$).

   Due to  (\ref{3.8})  we have
   \begin{equation}
   A_H - A_h = (H - h)[3 (S_2 - S_1^2) + 6 S_1 (H + h) - 6 (H^2 + Hh + h^2)],
    \label{3.20}
   \end{equation}
    and hence, by using  (\ref{3.3}),  (\ref{3.7}), we obtain
         \begin{equation}
                 Y_H - Y_h = (H-h)(mH + lh)[2 + 4 \alpha Q(H,h)],
                    \label{3.21}
           \end{equation}
         where
           \begin{equation}
           Q(H,h) = (m - 1)(m - 2)H^2 + 2 (m - 1)(l - 1) H h + (l - 1)(l - 2)h^2.
                \label{3.22}
           \end{equation}
         For $m > 1$ and $l > 1$ the quadratic form has the signature $(-,+)$.
         Due to $m H + lh \neq 0$ the set of eqs.  (\ref{3.19})
         is equivalent to another set of equations
           \begin{equation}
            E =0, \qquad   Y_H - Y_h = 0, \qquad  m H Y_H + l h Y_h = 0,
              \label{3.23}
           \end{equation}
          According to  (\ref{3.5a}) $E = 0$ implies
          $ h^i Y_i = m H Y_H + l h Y_h = 0 $
        and hence the third equation in  (\ref{3.23}) may be omitted.
        Using  restrictions (\ref{3.18}), relations (\ref{3.13a}) and (\ref{3.21}) we reduce the
        set eqs. (\ref{3.23}) to the following set of equations

         \begin{eqnarray}
           E =  m H^2 + l h^2 - (mH + lh)^2  + 2 \Lambda - \alpha [m (m-1) (m-2) (m - 3) H^4
           \nonumber \\
           + 4 m (m-1) (m-2) l H^3 h  + 6 m (m-1) l (l - 1) H^2 h^2
             \nonumber \\
           + 4 m l (l - 1) (l - 2) H h^3 + l (l - 1) (l - 2) (l - 3) h^4] = 0,
                \label{3.24}   \\
           1 + 2 \alpha Q(H,h) = 0,
           \label{3.25}
        \end{eqnarray}
      where $Q(H,h)$ is defined in (\ref{3.22}).
      \fnm[3]\fnt[3]{For general scheme of reduction see \cite{ChPavTop1}.}
      Thus, for the anisotropic solutions
      with two different Hubble parameters $H$ and $h$
      and non-static volume factor (see (\ref{3.13}) and (\ref{3.18})) the set $(n+1)$ polynomial
      eqs. of fourth order (\ref{3.5}) and (\ref{3.6}) is equivalent to the set of
      two eqs. (\ref{3.24}) and (\ref{3.25}) of fourth and second order respectively.

        \section{Stability of fixed point solutions $h^i(t) = v^i$}

        Here we   study the stability of static solutions $h^i(t) = v^i$ to eqs.
        (\ref{3.1}) and (\ref{3.2})  in linear approximation in pertubations. We put
        \begin{equation}
             h^i(t) = v^i +  \delta h^i(t),
             \label{4.l}
        \end{equation}
       $i = 1,\ldots, n$. By substitution (\ref{4.l}) into eqs.  (\ref{3.1}) and (\ref{3.2})
       we obtain in linear approximation  the following relations for perturbations $\delta h^i$
         \begin{eqnarray}
                    C_i(v) \delta h^i = 0, \label{4.2} \\
                    L_{ij}(v) \delta \dot{h}^j =  B_{ij}(v) \delta h^j,
                  \label{4.3}
           \end{eqnarray}
         where
           \begin{eqnarray}
         C_i =  C_i(v)  =  2 v_{i} - 4 \alpha G_{ijks}  v^j v^k v^s, \label{4.4} \\
         L_{ij} = L_{ij}(v) = 2 G_{ij} - 4 \alpha G_{ijks} v^k v^s,
                     \label{4.5} \\
          B_{ij} =  B_{ij}(v) = - (\sum_{k = 1}^n v^k)  L_{ij}(v) - L_i(v) + \frac{4}{3} v_{j}.
                     \label{4.6}
              \end{eqnarray}
        We remind that  $v_i = G_{ij} v^j$,
        $L_i(v) =  2 v_{i} - \frac{4}{3} \alpha  G_{ijks}  v^j v^k v^s$
         and $i,j,k,s = 1, \dots, n$.

          We put the following restriction on the matrix $L =(L_{ij}(v))$
          \begin{equation}
             ({\rm R }) \quad  \det (L_{ij}(v)) \neq 0,
                \label{4.7}
           \end{equation}
       i.e.  the matrix $L$  should be  invertible.

        Here we restrict ourselves by exponential solutions (\ref{3.4m}) with non-static volume
        factor, which is proportional to $\exp(\sum_{i = 1}^{n} v^i t)$,  i.e. we put
          \begin{equation}
            K = K(v) = \sum_{i = 1}^{n} v^i \neq 0.
                \label{4.8}
           \end{equation}
        Then we get from eq. (\ref{3.6})
        \begin{equation}
           L_i(v) = L_1 = \frac{2}{3} (\sum_{k = 1}^{n} v^k)^{-1}
            (G_{ij} v^i v^j - 4 \Lambda).
           \label{4.9}
         \end{equation}
      Due to  definition (\ref{3.3}) we have
      \begin{equation}
       \alpha A_i = \alpha  G_{ijks} v^j v^k v^s = \frac{3}{4} (2 v_i - L_1)
                 \label{4.10}
      \end{equation}
      and hence
            \begin{equation}
         C_i(v) = 2 v_i - 4 \alpha A_i = - 4 v_i + 3L_1.
         \label{4.11}
           \end{equation}

   We rewrite relation (\ref{4.6}) as
          \begin{equation}
         B_{ij} = - (\sum_{k = 1}^{n} v^k) L_{ij}(v) + \hat{B}_{ij},
         \qquad
         \hat{B}_{ij} =  - L_i(v) + \frac{4}{3}  v_{j}.
           \label{4.12}
           \end{equation}
    Due to $L_i(v) = L_1$ and (\ref{4.2}) we get
         \begin{equation}
         \hat{B}_{ij} \delta h^j = - L_1 \sum_{j =1}^n \delta h^j
          + \frac{4}{3} v_{j} \delta h^j = - \frac{1}{3} C_{j}(v) \delta h^j =0.
         \label{4.13}
         \end{equation}
     Hence  eq. (\ref{4.3}) reads
       \begin{equation}
     L_{ij}(v) \delta \dot{h}^j =  - (\sum_{k = 1}^{n} v^k) L_{ij} \delta h^j,
        \label{4.14}
       \end{equation}
  or, equivalently,
        \begin{equation}
       \delta \dot{h}^i = - (\sum_{k = 1}^{n} v^k)  \delta h^i,
        \label{4.15}
        \end{equation}
  $i = 1, \dots, n$. Here we used the restriction (\ref{4.7}).

 Thus, the set of linear equations on  perturbations (\ref{4.2}), (\ref{4.3}) is equivalent to the set
 of linear eqs. (\ref{4.2}), (\ref{4.15}), which has the following solution
   \begin{eqnarray}
       \delta h^i = A^i \exp( - K(v) t ),
       \label{4.16}  \\
         \sum_{i =1}^{n} C_i(v)  A^i =0.
         \label{4.16A}
   \end{eqnarray}
    $i = 1, \dots, n$. We remind that $K(v) = \sum_{k = 1}^{n} v^k$.

   Due to  (\ref{4.16}) that the following proposition is valid.

   {\bf Proposition 2.}
   {\em The fixed point solution $(h^i(t)) = (v^i)$ ($i = 1, \dots, n$; $n >3$)
   to eqs. (\ref{3.1}), (\ref{3.2}) obeying
   restrictions   (\ref{4.7}),  (\ref{4.8})
    is  stable under perturbations (\ref{4.l}) (as $t \to + \infty$)
   if  $K(v) = \sum_{k = 1}^{n} v^k > 0$ and
   it is unstable (as $t \to + \infty$) if $K(v) = \sum_{k = 1}^{n} v^k < 0$.}

  It follows from  (\ref{3.16}) that in the isotropic case the matrix (\ref{4.5}) reads
   \begin{equation}
   L_{ij} = \varphi(H) G_{ij}, \qquad \varphi(H) = 2 + 4 \alpha (n -2)(n-3)H^2.
    \label{4.17}
   \end{equation}
     Since the matrix   $(G_{ij}) = (\delta_{ij} - 1)$ is invertible
     (or, non-degenerate one) for $n > 1$
   (its inverse is  $(G^{ij}) = (\delta^{ij} - \frac{1}{n-1})$) then the
   matrix   $(L_{ij})$ is invertible if and only if $\varphi(H) \neq 0$.

    Now let us consider the matrix  (\ref{4.5}) for  the anisotropic case
   (\ref{3.13}) with two Hubble parameters obeying (\ref{3.18}).

        For the  the  ansatz   (\ref{3.13})  we obtain
      \begin{eqnarray}
      L_{\mu \nu} =  G_{\mu \nu} (2 + 4 \alpha S_{HH}),
            \label{4.18HH}   \\
       L_{\mu \alpha} = L_{\alpha \mu} = - 2 - 4 \alpha S_{Hh},
             \label{4.18Hh}   \\
       L_{\alpha \beta} = G_{\alpha \beta} (2 + 4 \alpha S_{hh}) .
             \label{4.18hh}
      \end{eqnarray}
     Here  $S_{HH}$, $S_{Hh}$ and  $S_{hh}$ are defined
     in  (\ref{3.17HH}), (\ref{3.17Hh}) and (\ref{3.17hh}), respectively.
      But here  we have  a remarkable coincidence (see (\ref{3.22}))
      \begin{equation}
      Q(H,h) = S_{Hh},
      \label{4.19}
      \end{equation}
     which implies $L_{\mu \alpha} = L_{\alpha \mu} = 0$  due to eq. (\ref{3.25}).
      Thus under restrictions   (\ref{3.18}) assumed the matrix
     $(L_{ij})$ has a block-diagonal form
      \begin{equation}
           (L_{ij}) = {\rm diag}(L_{\mu \nu}, L_{\alpha \beta} ).
              \label{4.20}
       \end{equation}
     This matrix is invertible if and only if  $m > 1$,  $l > 1$ and
       \begin{equation}
         S_{HH} \neq - \frac{1}{2 \alpha}, \qquad  S_{hh} \neq - \frac{1}{2 \alpha}.
         \label{4.21}
       \end{equation}
        We remind that  $m \times m$ matrix $(G_{\mu \nu})$ and $l \times l$  matrix
       $(G_{\alpha \beta})$ are invertible only for $m > 1$ and $l > 1$, respectively.

   {\bf Remark 2.} {\em   Recently, in ref. \cite{Pavl-15} a criterion  for stability
      of fixed point solutions in the model under consideration
      (and its extension to the Lovelock case)    was used.
      In our notations (see Introduction) it reads:
   \begin{equation}
    \frac{\partial \dot{h}^i }{\partial h^i} (v)
    = \frac{ \partial \varphi^i }{\partial h^i} (v) < 0,
     \label{4.22}
     \end{equation}
    $i = 1, \dots, n$.    It can be readily verified that for generic functions $f_0, f_i$ in eqs. (\ref{1.5}), (\ref{1.6})    the criterion (\ref{4.22}) is not  a necessary and/or a sufficient condition for the stability
    of the fixed point solutions.    Fortunately, for a special choice of  functions, e.g. for $f_0(h) = E(h)$,
    $f_i(\dot{h}, h) =   Y_i(\dot{h}, h) + \frac{1}{3} E(h ) = U_i(\dot{h}, h) $
    (see (\ref{3.2b}) and (\ref{4.13})), it  gives a correct result
     since  in this case
     \begin{equation}
      \frac{\partial \dot{h}^i }{\partial h^i} (v) =
     - \sum_{k = 1}^{n} v^k,
            \label{4.23}
      \end{equation}
      $i = 1, \dots, n$.
     Relation  (\ref{4.23}) is also valid for $f_i(\dot{h}, h) = \lambda U_i(\dot{h}, h)$       with $\lambda \neq 0$,  e.g. for the choice  $\lambda = -1$ used in \cite{Pavl-15}.
     We also note that in our notations $2\Lambda = \Lambda_P$, where $\Lambda_P$ is the $\Lambda$-term from ref. \cite{Pavl-15}.}

 \section{Examples}

 Here we consider several examples of exponential solutions and analyse their stability.

   \subsection{Isotropic solution}

     Let us consider the isotropic solution $v = (v^i) = (H,\ldots,H)$ to
     eqs. (\ref{3.5}), (\ref{3.6}) for $n > 3$.
     Due to $G_{ij}v^i v^j = n(1-n) H^2$ and (\ref{3.12}),
     eq. (\ref{3.5}) reads as follows
          \begin{equation}
         \label{5.1}
      2 F(H^2) = n(n-1) H^2 + \alpha n(n-1)(n-2)(n-3) H^4 = 2 \Lambda.
         \end{equation}
    Eq. (\ref{3.6}) is also equivalent to (\ref{5.1}) due to relation
     \begin{equation}
      \label{5.2}
     L_i = - 2(n-1) H + \frac{4}{3} \alpha (n-1)(n-2)(n-3) H^3,
     \end{equation}
    $i = 1, \dots, n$, which follows from   (\ref{3.3}),  (\ref{3.12}) and (\ref{3.7}).

    Let $\Lambda = 0$. The trivial solution $H =0$ is valid for any $\alpha$. This
    is the unique solution for $\alpha > 0$. For $\alpha < 0$ we have two non-trivial
    solutions \cite{Iv-09,Iv-10} with
      \begin{equation}
       \label{5.3}
         H^2 = \frac{1}{|\alpha| (n -2)(n -3)}.
      \end{equation}
      This solution was generalized in \cite{ChPavTop} to the case $\Lambda \neq 0$.

     Let us consider the case of generic $\Lambda$ in detail.
     First, we put  $\alpha > 0$.
      Then, a solution to eq. (\ref{5.1}) does exist if and only if  $\Lambda \geq 0$.
      For $\Lambda = 0$ we get $H = 0$, while for $\Lambda > 0$ we have two non-zero
      solutions for $H$ with $H^2 > 0$:
      \begin{equation}
     \label{5.4}
      H^2 = \frac{-n(n-1) + \sqrt{\Delta}}{2 \alpha  n (n-1)(n-2)(n-3)},
     \end{equation}
     where
      \begin{equation}
      \label{5.5}
          \Delta = n^2(n-1)^2 + 8 \Lambda \alpha  n(n-1)(n-2)(n-3).
      \end{equation}

       Now we put  $\alpha < 0$. A solution to eq. (\ref{5.1}) exists only if
      $\Lambda \leq \Lambda_{cr}$, where
      \begin{equation}
            \label{5.6}
              \Lambda_{cr} = - \frac{n(n-1)}{8  \alpha (n-2)(n-3)}
                \end{equation}
    is the maximum value of the function $F(H^2)$ from  (\ref{5.1}).
    For  $0 < \Lambda < \Lambda_{cr}$ (and $\alpha < 0$) we have two
    solutions for $H^2$ (or four solutions for $H$) which are given by relation
         \begin{equation}
         \label{5.7}
         H^2  = \frac{-n(n-1) \pm  \sqrt{\Delta}}{2 \alpha  n (n-1)(n-2)(n-3)}.
         \end{equation}
     For  $\Lambda = \Lambda_{cr}$ and $\alpha < 0$ we get one solution for $H^2$
     (or two solutions for $H$):
     \begin{equation}
      \label{5.8}
     H^2 =  H^2_{cr} = - \frac{1}{2 \alpha  (n-2)(n-3)}.
      \end{equation}
     The case $\Lambda = 0$ (and $\alpha < 0$) was mentioned above
     (two solutions for $H^2$, or three - for $H$).
    For  $\Lambda < 0$ (and $\alpha < 0$) we obtain one solution for $H^2$ (or two solutions for $H$):
     \begin{equation}
         \label{5.9}
          H^2 = \frac{-n(n-1) - \sqrt{\Delta}}{2 \alpha  n (n-1)(n-2)(n-3)}.
     \end{equation}

      Due to  (\ref{4.17})  the matrix $(L_{ij})$ is invertible for all solutions
      but $H = H_{cr}$  from (\ref{5.8}) for $\alpha < 0$,  since only in this case $\varphi(H) = 0$.
      The relation  $H = H_{cr}$ takes place only for  $\Lambda = \Lambda_{cr}$ and $\alpha < 0$ and hence   this case will be excluded from our analysis. Since $K(v) = n H$, the trivial solution $H=0$ for $\Lambda =0$ should  be also excluded from our consideration. It follows
      from Proposition 2 that all isotropic  solutions $v = (v^i) = (H,\ldots,H)$
      obeying  $H > 0$ and $H \neq H_{cr}$ for $\alpha < 0$ are stable while  all isotropic  solutions obeying  $H < 0$ and $H \neq H_{cr}$ for $\alpha < 0$ are unstable.

      Using (\ref{3.12}), (\ref{3.7}) and (\ref{4.4}) we obtain $C_i(v) = - (n-1)H \varphi(H) \neq 0$,
      $i = 1, \dots, n$, for  $H \neq 0$ and $H \neq H_{cr}$ for $\alpha < 0$.
      Under these restrictions on $H$,
      the solution for perturbations (\ref{4.16}), (\ref{4.16A}) reads as follows
           \begin{eqnarray}
             \delta h^i = A^i \exp( - n H t ),
             \label{5.10}   \\
               \sum_{i =1}^{n}   A^i =0,
               \label{5.10A}
         \end{eqnarray}
          $i = 1, \dots, n$. Relation (\ref{5.10}) was obtained earlier in \cite{Pavl-15}.

     \subsection{Anisotropic solutions with two Hubble parameters}

  In this subsection we  consider several examples of
  anisotropic solutions to eqs. (\ref{3.5}), (\ref{3.6})
  of the form  $v =(H, \ldots, H, h, \ldots, h)$,
  where $H$  the Hubble-like parameter corresponding  to $m$-dimensional isotropic subspace
  with $m \geq 3$ and $h$ is the Hubble-like parameter   corresponding
  to $l$-dimensional isotropic subspace with $l>1$.
  Here we put  $H > 0$.

  \subsubsection{Solution for $m =  3$, $l = 2$ and $\Lambda = 0$.}

 Let us consider the case $m =  3$, $l = 2$,  $\Lambda = 0$.
We have the following solution to the set of
 polynomial eqs. (\ref{3.24}), (\ref{3.25}) with $H > 0$:
  \begin{eqnarray}
   \label{5.2.1}
   H = \frac{1}{6} ( 7 + 4 \cdot 10^{1/3}  +  10^{2/3} )^{1/2} \alpha^{-1/2}
   \approx 0.750173 \alpha^{-1/2}, \\
   \label{5.2.2}
   h =  - \frac{1}{6} ( 7 - 0.5 \cdot10^{1/3}  +  10^{2/3} )^{1/2} \alpha^{-1/2}
   \approx - 0.541715 \alpha^{-1/2}.
  \end{eqnarray}

  It the approximate form  this  solution was found earlier in  \cite{Rat}, in
  analytic form  (different from (\ref{5.2.1}), (\ref{5.2.2})) it was  obtained in \cite{ChPavTop}.

  Using (\ref{3.17HH}) and (\ref{3.17hh}) we get
   \begin{equation}
    \label{5.2.3}
      S_{HH} =  2 h (2 H + h) \approx - 1.038610 \alpha^{-1}, \qquad
      S_{hh} =  6  H^2  \approx 3.376557 \alpha^{-1}.
    \end{equation}
  Relations (\ref{4.21}) are valid and hence the first restriction (\ref{4.7}) is satisfied.
  The second restriction (\ref{4.8}) is also satisfied since $K(v) = 3 H + 2 h > 0$. Thus,  due to Proposition 2, the solution is stable in agreement with \cite{Pavl-15}.

  \subsubsection{Solution for $m = l = 3$ and $\Lambda = 0$}

  Now we consider  solutions with $m =  3$, $l = 3$ and $\Lambda = 0$.
  There are two solutions to eqs. (\ref{3.24}), (\ref{3.25}) with $H > 0$:
 \begin{equation}
  \label{5.3.1}
   H_1 = \frac{1}{4}(\sqrt{5} - 1) \alpha^{-1/2},
   \qquad  h_1 =  \frac{1}{4}(- \sqrt{5} - 1) \alpha^{-1/2},
  \end{equation}
  and
  \begin{equation}
  \label{5.3.2}
   H_2 = \frac{1}{4}(\sqrt{5} + 1) \alpha^{-1/2},
   \qquad h_2 =  \frac{1}{4}(- \sqrt{5} + 1) \alpha^{-1/2}.
  \end{equation}

 For the first solution  we get
   \begin{equation}
   \label{5.3.3}
  S_{HH} =  \frac{3}{4} (\sqrt{5} + 1) \alpha^{-1},
  \qquad  S_{hh} = \frac{3}{4} (- \sqrt{5} + 1) \alpha^{-1},
   \end{equation}
 while  for the second one we obtain
    \begin{equation}
    \label{5.3.4}
     S_{HH} =    \frac{3}{4} (- \sqrt{5} + 1) \alpha^{-1},
     \qquad  S_{hh} = \frac{3}{4} (\sqrt{5} + 1) \alpha^{-1}.
   \end{equation}
  In both cases  relations (\ref{4.21}) are satisfied and
  hence the first restriction (\ref{4.7}) is valid.
   The second restriction (\ref{4.8}) is also valid for any of
   these solutions  since $K(v_1) = 3 H_1 + 3 h_1 = - \frac{3}{2} \alpha^{-1/2} < 0$ and
   $K(v_2) = 3 H_2 + 3 h_2 = \frac{3}{2} \alpha^{-1/2} > 0$. According to Proposition 2
   the first solution (\ref{5.3.1}) is unstable,
    while the second one (\ref{5.3.2}) is stable.

  \subsubsection{Solution for $m=11$,  $l=16$ and $\Lambda = 0$}

       For $\Lambda = 0$  the solution (\ref{3.4m})
       with  $v=(v^i)$ from (\ref{3.13}), $m=11$, $l=16$ and
        \begin{equation}
          \label{5.4.1}
           H= \frac{1}{\sqrt{15 \alpha}}, \qquad  h= -\frac{1}{2 \sqrt{15 \alpha}}
          \end{equation}
    was found in  \cite{IvKob}.  This solution describes the zero variation
    of the effective cosmological constant $G$.

      The calculations give us
         \begin{equation}
              S_{H H}= - \frac{4}{5} \alpha^{-1},  \quad S_{hh} =  \frac{1}{10} \alpha^{-1}.
              \label{5.4.2}
         \end{equation}
        Due to  (\ref{4.21})  the  symmetric matrix $(L_{ij})$, which has a block-diagonal form,
        is invertible, i.e. the condition  (\ref{4.7}) is satisfied.

         Using (\ref{4.9}) and (\ref{4.11}) we find $(C_i) = (C_{\mu} = 12 H, C_{\alpha} = 18H)$.
        From (\ref{4.16}) we get the following solution for perturbations
        \begin{eqnarray}
                 \delta h^i = A^{i} \exp( - 3 H t), \label{5.17} \\
                 2 \sum_{\mu = 1}^{11} A^{\mu} + 3  \sum_{\alpha = 12}^{27} A^{\alpha} = 0,
                \label{5.4.3}
            \end{eqnarray}
         where $H= \frac{1}{   \sqrt{15 \alpha}}$, $i = 1, \dots, 27$.
         Thus, the solution (\ref{5.4.1}) is stable, as
           $t \to + \infty$.

         \subsubsection{Solution for $m=15$,  $l=6$ and $\Lambda = 0$}

         Now we consider another exponential solution   (\ref{3.4m}) from  \cite{IvKob}
         with  $v=(v^i)$ from (\ref{3.13}), $m=15$,  $l=6$,  $\Lambda = 0$
         and
      \begin{equation}
           \label{5.5.1}
            H=\frac{1}{6} \alpha^{-1/2}, \qquad  h= - \frac{1}{3} \alpha^{-1/2}.
      \end{equation}

          We get
      \begin{equation}
               S_{HH}= - \alpha^{-1}, \qquad S_{hh} =  \frac{1}{2} \alpha^{-1}.
               \label{5.5.2}
      \end{equation}
  According to (\ref{4.21})  the symmetric block-diagonal matrix $(L_{ij})$  is non-degenerate one.

  By using (\ref{4.9}) and (\ref{4.11}) we get
    $(C_i) = (C_{\mu} = \frac{14}{3}, C_{\alpha} = \frac{20}{3})$.
  Due to  (\ref{4.16}) the solution for perturbations reads
        \begin{eqnarray}
          \delta h^i = A^{i} \exp( - 3 H t) = A^{i} \exp( - \frac{1}{2} \alpha^{-1/2} t), \label{5.17a} \\
            7 \sum_{\mu = 1}^{15} A^{\mu} + 10  \sum_{\alpha = 16}^{21} A^{\alpha} = 0,
           \label{5.5.3}
        \end{eqnarray}
          $i = 1, \dots, 21$.
   Hence, the solution (\ref{5.5.1}) is  stable as $t \to + \infty$.

   {\bf Remark 3}. {\em The stability of this solution as well as the previous one
   was also proved   in ref. \cite{EIK} by using  rather tedious
   calculations based on   relations (\ref{4.3}) and  (\ref{4.6})
   without using the identity (\ref{4.13}).}

  \subsubsection{Solutions with $m  \geq 3$, $l > 1$ and certain $\Lambda > 0$}

  Here we consider the following solution  to eqs. (\ref{3.24}), (\ref{3.25}) for $m > 2$,
   $l > 1$ and  $\alpha < 0$:
    \begin{equation}
    \label{5.6.l}
     H^2 = - \frac{1}{2 \alpha (m-1)(m-2)}, \qquad   h = 0,
    \end{equation}
  which is valid for
     \begin{equation}
     \label{5.6.2}
        \Lambda = - \frac{m (m +1)}{8 \alpha (m-1)(m-2)} > 0.
     \end{equation}

  We get  from (\ref{3.17HH}) and (\ref{3.17hh})
    \begin{equation}
       \label{5.6.3}
        S_{HH} =  (m-2)(m -3) H^2 =  - \frac{m -3}{2 \alpha (m-1)}  \neq - \frac{1}{2 \alpha}
     \end{equation}
  and
    \begin{equation}
      \label{5.6.4}
       S_{hh} =  m (m-1) H^2 =  - \frac{m}{2 \alpha (m-2)}  \neq - \frac{1}{2 \alpha},
    \end{equation}
  which implies the fulfilment of the restriction (\ref{4.7}) (here $m >2$ and $l > 1$). Since
  $K(v) = m H$ we get from  Proposition 2 that the cosmological solution (\ref{3.4m})
  with $H$, $h$ from   (\ref{5.6.l}) is stable for $H > 0$ and unstable for $H < 0$.

 \subsection{A subclass of solutions with zero variation of $G$}

 The 4d effective gravitational constant is proportional to inverse volume scale factor
 of the internal space (see  \cite{BIM,Mel,IvMel-14}), i.e.
    \begin{equation}
   \label{6.G0}
     G \sim \prod_{i=4}^{n} [a_i(t)]^{-1} ,
    \end{equation}
   where  $a_i(t) = \exp(\beta^i(t))$.

  {\bf Remark 4}. {\em Here $G = G_{eff}^{J}(t)$ is four-dimensional effective
  gravitational constant which appear in (multidimensional analogue of) the so-called
  Brans-Dicke-Jordan (or simply Jordan) frame  \cite{RZ-98}. In this case the physical 4-dimensional
  metric  $g^{(4)}$ is defined as  4-dimensional section of the multidimensional metric $g$, i.e.
   $g^{(4)} = g^{(4,J)}$, where  $g = g^{(4,J)} +  \sum_{i=4}^{n} a^{2}_i (t)  dy^i \otimes dy^i$.
   When  the Einstein-Pauli (or simply Einstein) frame  is used, we put   $g^{(4)} = g^{(4,E)} = (\prod_{i=4}^{n} a_i (t)) g^{(4,J)}$ \cite{RZ-98,I-96} and hence we get the effective
   gravitational constant  to be an exact constant:
   $G_{eff}^{E} = G_{eff}^{J}(t) \prod_{i=4}^{n} a_i(t) = {\rm const} $  \cite{RZ-98}}.

  For the  solutions (\ref{3.4m})   we obtain the following relations
  \begin{equation}
     \label{6.G}
  G(t) = G(0) \exp{(- K_{int} t) },  \quad  K_{int}(v) = \sum_{i=4}^{n} v^i,
   \end{equation}
 which imply
  \begin{equation}
  \label{6.Gd}
 \frac{\dot{G}}{G} =  - K_{int}(v).
  \end{equation}

 Now, let us consider a subclass of cosmological solutions (\ref{3.4m})
 which obey restriction (\ref{4.7}) and describe  an exponential isotropic expansion of
3-dimensional flat factor-space with $v^1 = v^2 =v^3 = H >0$ with
 zero variation of $G$. Then we get from  (\ref{6.Gd})  $K_{int}(v) = 0$ and
hence $K(v) = \sum_{i=1}^{n} v^i = 3H + K_{int}(v) = 3H > 0$. According to Proposition 2
any solution from this subclass is stable. Three solutions from the previous subsection:
(\ref{5.4.1}), (\ref{5.5.1}) and (\ref{5.6.l}) with $m =3$ (and $l > 1$) belong to this
subclass.

{\bf Remark 5}. {\em It should be noted that for $K(v) = 0$ and $v^1 = v^2 =v^3 = H >0$ we obtain
$K_{int}(v) = - 3H$ and hence $\frac{\dot{G}}{G} =  3H > 0$.}

 \section{Conclusions}

 We have considered the  $(n+1)$-dimensional  Einstein-Gauss-Bonnet (EGB) model
 with the $\Lambda$-term.  By  using the  ansatz with diagonal  cosmological  metrics,
 we have studyed the stability of solutions with exponential dependence of scale factors
 $a_i \sim \exp{ ( v^i t) }$, $i =1, \dots, n $,
 with respect to synchronous time variable $t$ in dimension $D > 4$.

 The problem was reduced to the analysis of stability of the fixed point solutions
 $h^i(t) = v^i$ to  eqs. (\ref{3.1}) and (\ref{3.2a}),
 where $h^i(t)$ are  Hubble-like parameters.

 In this paper a set of equations for perturbations $\delta h^i$  was considered
 (in linear approximation) and  general solution to these equations was found.
 We have proved (in Proposition 2) that the solutions  with non-static volume factor, i.e. with
 $K(v) =  \sum_{k = 1}^{n} v^k \neq 0$, which  obey restriction  (\ref{4.7}),
   are stable if  $K(v) > 0$  while they are unstable if $K(v) < 0$.

We have also proved (in  Proposition 1) that for any  exponential solution
with $v = (v^1,...,v^n)$  there are no more than  three different  numbers among
 $v^1,...,v^n$, if  $\sum_{i=1}^n v^i  \neq 0$.

Here we have presented several examples of stable cosmological solutions with exponential
behavior of scale factors.  Among them the isotropic solution $v =(H, \ldots, H)$ and
several  anisotropic solutions with two Hubble parameters
$v =(H, \ldots, H, h, \ldots, h)$ were  considered.
The isotropic solution is stable if $H > 0$ and $H \neq H_{cr}$ for $\alpha < 0$
(see (\ref{5.8})).  For the anisotropic case our examples deal
 with  the Hubble-like parameter $H > 0$  corresponding  to $m$-dimensional flat subspace
 with $m \geq 3$ and  the Hubble-like parameter $h$  corresponding
 to $l$-dimensional flat subspace with $l > 1$.
 This  subclass of (anisotropic) solutions contains the following cases:
 i) $m =  3$, $l = 2$, $\Lambda = 0$; ii) $m = l = 3$, $\Lambda = 0$;
 iii) $m=11$,  $l=16$, $\Lambda = 0$; iv) $m=15$,  $l=6$,  $\Lambda = 0$;
 v) $m  \geq 3$, $l > 1$, $\Lambda > 0$.
 We have also shown that general
 solutions with   $v^1 = v^2 =v^3 = H > 0$ and   zero variation of
 the  effective gravitational constant  are stable if the restriction (\ref{4.7}) is obeyed.

 \begin{center}
 {\bf Acknowledgments}
 \end{center}

This paper was funded by the Ministry of Education and Science of the Russian Federation in the Program to increase the competitiveness
of Peoples' Friendship University (RUDN University) among the world's leading research and education centers in the 2016-2020 and  by the  Russian Foundation for Basic Research,  grant  Nr. 16-02-00602.

\end{document}